\documentclass[prb,aps,twocolumn,draft,tightenlines,%
showpacs,floatfix]{revtex4}
\usepackage{psfig}
\usepackage{amssymb}
\usepackage{colordvi}
\begin{document}
\title{Diffusion and transport of spin pulses in an $n$-type
 semiconductor quantum  well}

\author{L. Jiang}
\affiliation{Hefei National Laboratory for Physical Sciences at
  Microscale,
  University of Science and Technology of China, Hefei,
  Anhui, 230026, China}
\affiliation{Department of Physics, University of Science and Technology of
  China, Hefei, Anhui, 230026, China}
\altaffiliation{Mailing Address.}

\author{M. Q. Weng}
\affiliation{Hefei National Laboratory for Physical Sciences at
  Microscale,
  University of Science and Technology of China, Hefei,
  Anhui, 230026, China}
\affiliation{Department of Physics, University of Science and Technology of
  China, Hefei, Anhui, 230026, China}
\author{M. W. Wu}
\thanks{Author to whom all correspondence should be addressed}
\email{mwwu@ustc.edu.cn.}
\affiliation{Hefei National Laboratory for Physical Sciences at
  Microscale,
  University of Science and Technology of China, Hefei,
  Anhui, 230026, China}
\affiliation{Department of Physics, University of Science and Technology
of China, Hefei, Anhui, 230026, China}

\author{J. L. Cheng}
\affiliation{Department of Physics, University of Science and Technology
of China, Hefei, Anhui, 230026, China}

\date{\today}

\bigskip

\begin{abstract}
We perform a theoretical investigation on the time evolution of
spin pulses in an $n$-type GaAs (001) quantum well  with and without
external electric field at high temperatures
by constructing and numerically solving
the kinetic spin Bloch equations and the Poisson equation,
with the electron-phonon, electron-impurity and electron-electron Coulomb
scattering explicitly included. The effect of the Coulomb scattering,
especially the effect of the Coulomb drag on the spin diffusion/transport
is investigated and it is shown that the
spin oscillations and spin polarization reverse along the direction
of spin diffusion in the absence of the applied magnetic field,
which were originally predicted in the absence of the Coulomb scattering
by Weng and Wu [J. Appl. Phys. {\bf 93}, 410 (2003)],
can sustain the Coulomb scattering at high temperatures ($\sim 200$\ K).
The results obtained are consistent
with a recent experiment in bulk GaAs but at a very low
temperature (4\ K) by Crooker and Smith [Phys. Rev. Lett. {\bf 94}, 236601
(2005)].

\end{abstract}
\maketitle

Semiconductor spintronics,  which aims at replacing the charge
degrees of freedom  with the spin ones in electronic
devices, has attracted substantial  attention
recently.\cite{wolf,spintronics,das} A thorough understanding of
transport of spin polarized electrons from one place to another
by means of electrical or diffusive current is an important prerequisite
for the realization of spintronic devices such as spin transistors and
spin valves. Study of the time evolution of a spin
pulse provides an ideal  platform for this purpose and has been
investigated by two of us based on kinetic spin Bloch
equations in $n$-type GaAs quantum wells (QW's).\cite{weng_JAP, weng}
It is pointed out there that for Zinc-blende semiconductors,
the spin coherence
 $\rho_{{\bf k}\sigma-\sigma}=\langle
c^\dagger_{{\bf k}\sigma}c_{{\bf k}-\sigma}\rangle$
 play an important role in the spin diffusion/transport.
It leads to some striking behaviors, such as at high
temperatures (200\ K) the spin
polarization can be opposite to the initial one along the spin diffusion
 even in the absence of an external magnetic field.\cite{weng_JAP, weng}
This spin oscillation along the direction of spin diffusion
can not be obtained under the framework of
quasi-independent electron model.\cite{qusi}
This behavior has been reproduced later by
Monte Carlo simulations by Pershin.\cite{pershin}
Very recently Crooker and Smith have also shown experimentally\cite{crooker}
this feature at very low temperature (4\ K)
in an $n$-type bulk GaAs with strains which provide a
similar effect as the Dresselhaus term\cite{dress} in GaAs QW's in
our investigation.

However, it is noted that in our previous spin transport
studies\cite{weng_JAP,weng} as well as the Monte Carlo
simulations,\cite{pershin} the electron-electron Coulomb
scattering is not included. Studies of spin kinetics have
shown that the Coulomb scattering plays a crucial role in
spin dephasing and relaxation.\cite{wu_epjb_2000,glazov_2002,many,weng2}
Therefore, it should also strongly affect the spin coherence in
spin diffusion and transport. In addition, the diffusion of a
pure spin pulse is formed as the spin-up and -down
electrons  moving in opposite directions. This causes
inevitably a Coulomb drag\cite{drag,drag1} between
electrons of opposite spins and
therefore may strongly alter the spin diffusion/transport.
All these considerations inspire us to investigate how the
Coulomb scattering can affect the spin diffusion/transport.
Moreover, inclusion of the Coulomb
scattering further allows us  to investigate the transport
properties of high spin polarization.
To our knowledge, a theoretical investigation of spin diffusion/transport
with the Coulomb scattering explicitly
included has never been performed in the literature.

We start our investigation in an $n$-type (001) GaAs QW
along the $z$-axis of small well
width $a$.  The dominant spin dephasing mechanism here is the
D'yakonov and Perel' (DP) mechanism.\cite{dp, meier}
By taking account of the DP term, the spin kinetic Bloch
equations can be written as\cite{weng_JAP}
\begin{eqnarray}
\label{BEQ}
&&\frac{\partial \rho(\mathbf{R}, \mathbf{k}, t)}{\partial
    t}-\frac{1}{2}\{\mathbf{\bigtriangledown}_{\mathbf{R}}
  \overline{\varepsilon}(\mathbf{R}, \mathbf{k}, t),
  \mathbf{\bigtriangledown}_{\mathbf{k}}\rho(\mathbf{R}, \mathbf{k},
  t)\} \nonumber\\
  && \hspace{0.7cm}+{\frac{1}{2}}\{\mathbf{\bigtriangledown}_{\mathbf{k}}\overline
     {\varepsilon}(\mathbf{R},\mathbf{k},t),\mathbf{\bigtriangledown}_
     {\mathbf{R}}\rho(\mathbf{R},\mathbf{k},t)\}
     -\left.\frac{\partial\rho(\mathbf{R},
    \mathbf{k},t)}{\partial t}\right|_{\mathtt{c}} \nonumber\\
  && \hspace{0.7cm} =\left.\frac{\partial\rho(\mathbf{R},
\mathbf{k}, t)}{\partial t}\right|_{\mathtt{s}}
\end{eqnarray}
with $\rho(\mathbf{R},\mathbf{k},t)$ standing for the
single-particle density matrix. The diagonal elements describe the
electron distribution functions
$\rho_{\sigma\sigma}(\mathbf{R},\mathbf{k},t)=
f_{\sigma}(\mathbf{R},\mathbf{k},t)$ of wave vector
$\mathbf{k}=(k_x, k_y)$ and spin $\sigma$ ($=\pm 1/2$) at position
${\mathbf{R}}$ and time $t$. The off-diagonal elements
$\rho_{\sigma - \sigma}({\bf R},{\bf k},t)$ describe the
inter-spin-band correlations (coherence) for the spin coherence.
$\overline{\varepsilon}(\mathbf{R},\mathbf{k},t)=
\varepsilon_{\mathtt{k}}\delta_{\sigma\sigma^{\prime}}+
\mathbf{h}(\mathbf{k})\cdot
{\mathbf{\mbox{\boldmath$\sigma$\unboldmath}}_{\sigma\sigma^{\prime}}}/{2}
-e\psi(\mathbf{R},t)+\Sigma_{\sigma\sigma^{\prime}}(\mathbf{R},
\mathbf{k},t)$. Here $\varepsilon_{k}=\mathbf{k}^{2}/2m^{\ast}$ is
the energy spectrum with $m^{\ast}$ denoting the electron
effective mass. $\mbox{\boldmath$\sigma$\unboldmath}$ are the
Pauli matrices. $\mathbf{h}(\mathbf{k})$ denotes the effective
magnetic field from the DP term which contains contributions from
both the Dresselhaus term\cite{dress} and the Rashba
term.\cite{ras} For GaAs QW, the leading term is the Dresselhaus
term, which can be written as\cite{eppenga} $h_x(\mathbf{k}) =
\gamma k_x (k_y^2-\pi^2/a^2)$ and $h_y(\mathbf{k}) = \gamma k_y
(\pi^2/a^2-k_x^2)$ with $\gamma$ standing for the spin-orbit
coupling strength.\cite{aronov} The electric potential
$\psi(\mathbf{R},t)$ satisfies the Poisson equation\cite{poisson}
\begin{equation}
\label{poi}
\mathbf{\bigtriangledown}_{\mathbf{R}}^{2}\psi(\mathbf{R},
t)=e[n(\mathbf{R}, t)-n_{0}(\mathbf{R})]/(a\epsilon_{0})\ ,
\end{equation}
in which $n(\mathbf{R},t)=\sum_{\sigma\mathbf{k}}f_{\sigma}(\mathbf{R},
\mathbf{k}, t)$ is the electron density at position $\bf R$ and time $t$
and $n_{0}(\mathbf{R})$ represents the positive background electric charge
density. $\Sigma_{\sigma\sigma^{\prime}}(\mathbf{R},\mathbf{k}, t) =
-\sum_{\mathbf{q}}V_{\mathbf{q}}\rho_{\sigma\sigma^{\prime}}(\mathbf{R},
\mathbf{k}-\mathbf{q}, t)$ is the Coulomb Hartree-Fock self-energy.
 $\left.\partial \rho(\mathbf{R},\mathbf{k}, t)/{\partial t}\right
|_{\mathtt{c}}$
and $\left.\partial \rho(\mathbf{R}, \mathbf{k}, t)/{\partial
 t}\right|_{\mathtt{s}}$ in Eq.\ (\ref{BEQ})
are the coherent and scattering terms respectively. Their expressions
are given in detail in Ref.\ \onlinecite{weng2}.
For the scattering, we include the electron-electron Coulomb scattering,
the electron-phonon scattering and the
electron--non-magnetic-impurity scattering. For electron-phonon
scattering, only the electron--longitudinal optical (LO) phonon
scattering is considered as we concentrate on the  high
temperature regime ($200$~K) where the electron-acoustic phonon
scattering is negligible. As we focus on
narrow QW's, the separation of the subband is
large enough so that only the lowest subband is needed in our calculation.

We assume that at initial time $t=0$ there is a pure spin pulse
polarized along $z$-axis and centered at $x=0$.
The electrons are
locally in Fermi distribution, {\em i.e.},
$f_{\sigma}(x,\mathbf{k},0)=(\mathtt{exp}\{[\epsilon_{\mathbf{k}}
-\mu_{\sigma}(x)]/T_{e}\}+1)^{-1}$, where $\mu_\sigma(x)$ stands for the
chemical potential of electrons with spin $\sigma$ at position $x$ and
is determined by the corresponding electron density:
$N_{\sigma}(x,0)=\sum_{\mathbf{k}}f_{\sigma}(x,\mathbf{k},0)$. The
shape of the initial spin pulse is assumed to be Gaussian like
\begin{equation}
\Delta N_{\sigma}(x, 0)=N_{1/2}-N_{-1/2}
=\Delta N_{0}\mathtt{e}^{-x^{2}/\delta x^{2}}
\end{equation}
with the $\Delta N_{0}$ and the $\delta x$ representing the peak and the
width of the spin pulse respectively. We further assume that there
is no spin coherence at the initial time, $\rho(x, \mathbf{k}, 0)=0$.

We numerically solve the kinetic Bloch equations (\ref{BEQ})
together with the Poisson equation (\ref{poi}) following the
method developed by two of us in Ref.\ \onlinecite{weng_JAP}.
The tricks of how to deal with the scattering, especially the
Coulomb scattering, are laid out in detail in  Ref.\ \onlinecite{weng2}.
It is due to these tricks which allow us to include the Coulomb scattering
in the spin diffusion/transport computation.
By solving the equations,
the temporal evolution of the electron
distribution functions $f_{\sigma}(\mathbf{R},\mathbf{k},t)$ and the
spin coherence $\rho(\mathbf{R}, \mathbf{k}, t)$ at different positions
$\mathbf{R}$  are obtained self-consistently.
From these quantities, all the transport information, such as
mobility, particle and spin diffusion lengths as well as spin dephasing and
relaxation are obtained explicitly
 {\em without} any fitting
parameters and approximations such as the relaxation
time approximation wildly used in the literature.
Moreover, as we include the Coulomb scattering explicitly, we
are able to calculate the situation with high spin polarization
and the situation with high external electric field in spin transport,
where the Coulomb scattering is crucial in the relaxation and
thermalization of electrons. It is noted that
both cases have been studied in a much simplified case, {\em i.e.},
pure spin precessions without diffusion/transport (in the absence of
the spacial degree of freedom).\cite{weng2,wu}

\begin{figure}[htbp]
  \centering
  \psfig{file=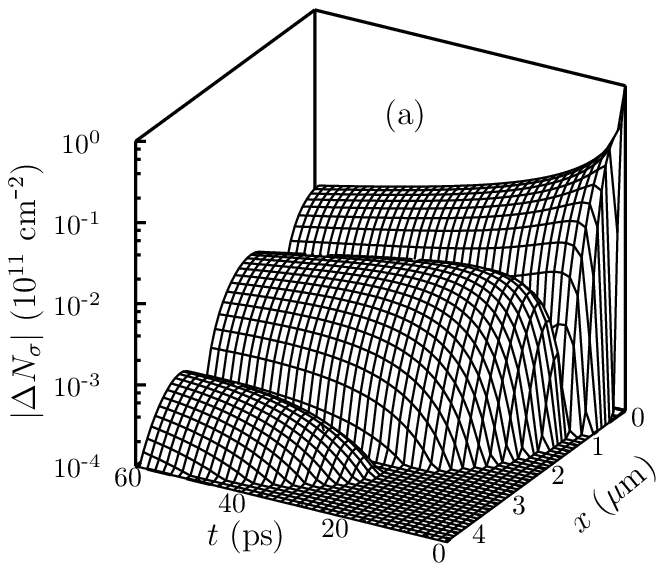,width=6.cm}
  \psfig{file=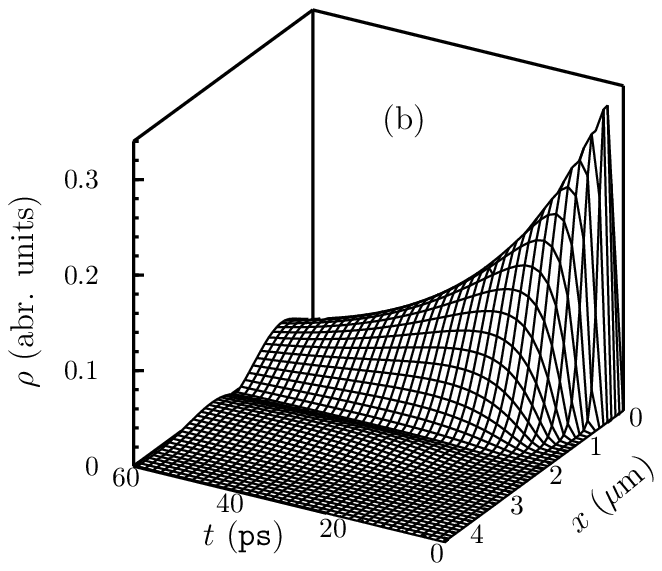,width=6.cm}
  \caption{The absolute value of the spin imbalance $|\Delta N_{\sigma}|$ as
    well as the incoherently summed spin coherence $\rho$ {\em
      v.s.} the position $x$ and the time $t$ for the spin pulse
with  $\Delta N_{0}=10^{11}$\
      cm$^{-2}$ and $\delta x =
    0.1$\ $\mu$m. Both the electron-LO phonon scattering and
  the  electron-electron Coulomb scattering are included.}
  \label{fig1}
\end{figure}

\begin{figure}[htbp]
  \centering
  \psfig{file=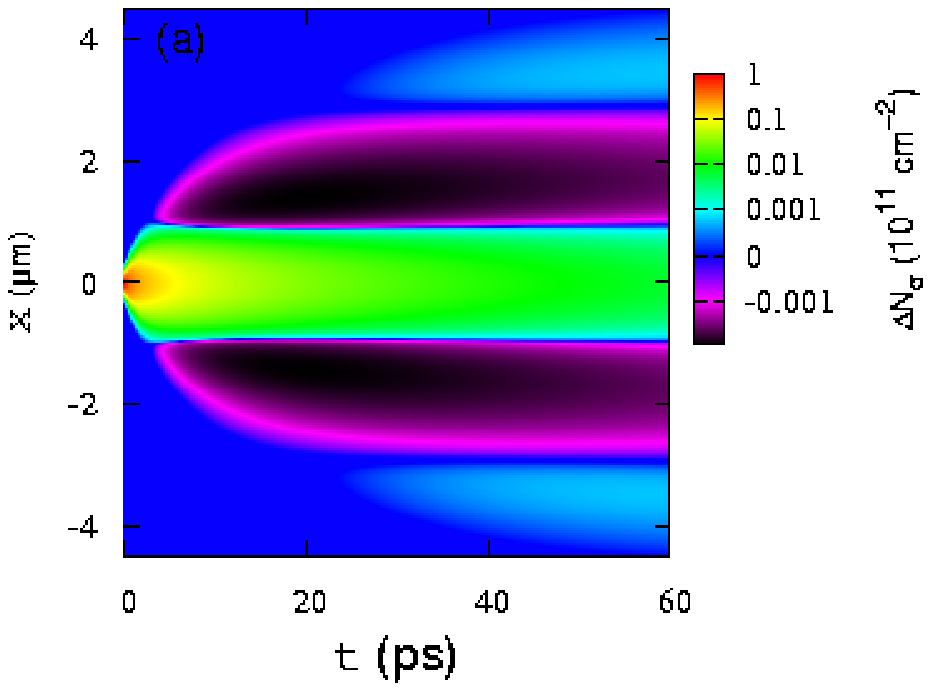,width=8.cm}
  \psfig{file=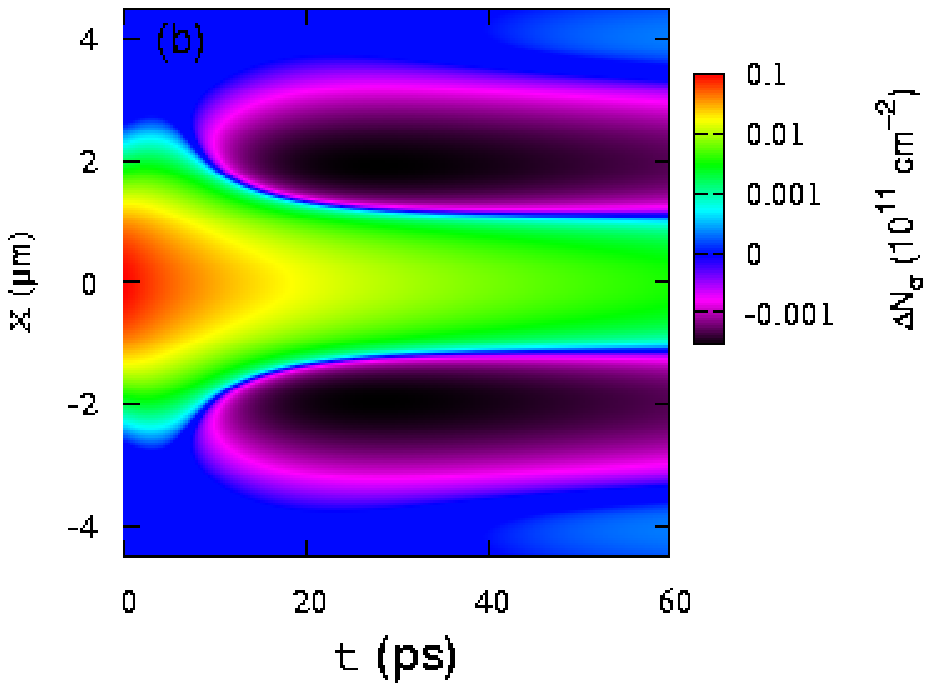,width=8.cm}
  \caption{(Color online) The contour plots of spin imbalance $\Delta
    N$ {\em v.s.}
  the  position $x$ and the time $t$ for different pulse widths. ($a$):
    $\delta x = 0.1~\mu$m; ($b$): $\delta x = 1.0~\mu$m. $T=200$\ K.}
\label{fig2}
\end{figure}

\begin{figure}[htbp]
\psfig{file=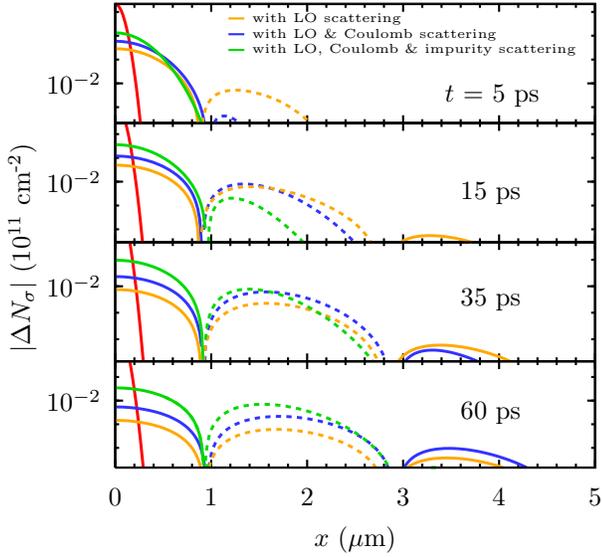,width=8.cm} \caption{(Color online) The
absolute values of the spin imbalance
  $|\Delta N_{\sigma}|$
{\em v.s.}  the position $x$ at certain times $t$ for the cases
with only the electron-LO phonon scattering (orange); the
electron-phonon and electron-electron scattering (blue); and
 the electron-phonon, electron-electron and electron-impurity scattering
(green). the red curve shows the initial spin pulse.
 $\Delta N_{0}=10^{11}$\ cm$^{-2}$, $\delta x=0.1$\
    $\mu$m and $N_i=0.5N_e$. Solid curves:
  $\Delta N > 0$; Dashed curves: $\Delta
  N < 0$. }
\label{fig3}
\end{figure}

\begin{figure}[htbp]
\psfig{file=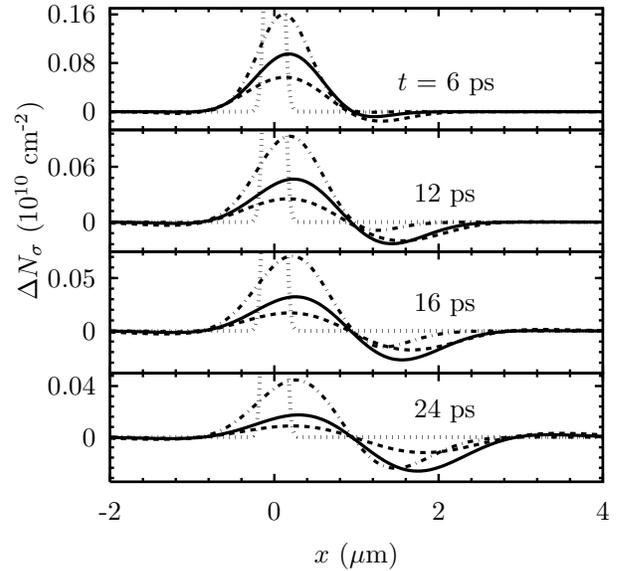,width=8.cm} \caption{The spin imbalance
$\Delta N_{\sigma}$ {\em v.s.} the position $x$ at certain times
$t$. Dotted curve: the initial spin pulse; Dashed curve: with only
electron-LO phonon scattering; Solid curve: with electron-LO
phonon and electron-electron Coulomb scattering;
  Chain curve: with electron-LO phonon, electron-impurity and electron-electron
Coulomb scattering. $\Delta N_{0}=10^{11}$\ cm$^{-2}$, $\delta
x=0.1$\ $\mu$m and $N_i=0.5N_e$.} \label{fig4}
\end{figure}

We study the temporal
evolution of the spin pulse at $T=200$\ K. The well
width and the total electron density are chosen to be $7.5$\ nm and
$4\times 10^{11}$\ cm$^{-2}$ respectively throughout our calculation.
The material parameters of GaAs are listed in Ref.\ \onlinecite{weng2}.
The main results of our calculation are summarized
in Figs.\ \ref{fig1}-\ref{fig4}.

We first examine if the Coulomb scattering can change the
spin oscillations along the direction of the spin diffusion.
 In Fig.\ 1 the absolute value
of the spin imbalance $|\Delta N_{\sigma}(x, t)|$ for the small pulse
width of $\delta x = 0.1$\ $\mu$m as well as the incoherently-summed
spin coherence $\rho(x,t)=\sum_{\mathbf{k}}|\rho_{\mathbf{k}\sigma
  -\sigma}(x,t)|$ are plotted as functions of position $x$
along the diffusion direction and
the time $t$ in the absence of the electric field.
Both the electron-LO phonon scattering and
the electron-electron Coulomb scattering
are included in the calculation. From the figure, it can be seen that
the spin signal around the center $x < 0.06$\ $\mu$m decays very fast
due to the diffusion as well as due to the spin dephasing. For the region of
0.06\ $\mu$m $< x <$ 0.9\ $\mu$m, due to the strong diffusion from the
center, the spin signal first amplifies then decays resulting from the
weakening of the diffusion as well as the dephasing.
For the regions further away from the pulse center, {\em i.e.}, 0.9\ $\mu$m$<x
<1.02$\ $\mu$m and 2.9\ $\mu$m$ < x <$ 2.94\ $\mu$m, we find
that the above counter effects
result in the oscillations of the spin polarization
with time, which were first discovered in the
absence of the electron-electron scattering.\cite{weng_JAP,weng}
Furthermore  it is noted that the striking
feature  that the spin polarization can be
opposite to the initial one  in
the absence of a magnetic field is still retained
in the presence of  the Coulomb
scattering. It is seen from the figure that the spin polarization
first reverses at the time about 3\ ps and the position around
1.02\ $\mu$m. And for the region  1.02\ $\mu$m $< x < 2.96$\ $\mu$m,
one observes
a second peak with its polarization opposite to the initial one.
And then after  16\ ps, the spin polarization reverses
back to the initial one in the region $x > 2.96$\ $\mu$m.

We next investigate the effect of pulse width on the
spin diffusion by comparing two pure spin pulses with different
pulse widths ($\delta x=0.1$\ $\mu$m and 1\ $\mu$m) but the same
total spin imbalance ($\int\Delta
  N_{\sigma}(x, t=0) dx=1.8 \times 10^{10}$\  cm$^{-1}$). The spin
polarizations at the center of the spin pulse is $25\%$ and $2.5\%$
respectively. The spin diffusions of the two spin pulses are
plotted against $t$ and $x$ in counter plots in
Fig ~\ref{fig2}. In the calculation we include the electron-electron and
electron-LO phonon scattering. It is seen from the figure that
the decay at the center of the narrow spin pulse
is much faster than that of the wide one and that
it is much faster for the narrow spin signal to diffuse out.
Also the diffusion of the narrow spin pulse
shows  stronger spin oscillations along the direction of the
diffusion. The reverse of the spin polarization occurs far away from the
position of the initial spin pulse. Moreover, the spin polarization
reverses back to the initial direction for the narrow spin pulse
at positions farther away from the initial pulse. This peak can hardly
be seen for the case of the wide pulse.
These features are understood as following:
The narrower spin pulse leads to larger gradients of the spin
densities and hence a larger broadening in the $k$-space,
and then leads to a faster diffusion.
Moreover, a larger momentum $k$ indicates a stronger
effective magnetic field from the DP term, which results in a
stronger spin oscillations.

In Fig.\ \ref{fig3} we investigate the effect of the Coulomb
drag\cite{drag,drag1}
on the spin diffusion by plotting the absolute values  of the
spin imbalance as functions of $x$ at $t=5$, 15, 35 and 60\ ps,
with and without the Coulomb scattering.
Contrary to the common impression that the Coulomb scattering
conserves the total  momentum and should not change the spin diffusion
dramatically, one finds that when the Coulomb scattering
is included, both the diffusion of the spin pulse and
the decay of the spin polarization at the center of the pulse
become much slower. Specifically at $t=15$\ ps, a third
peak with the spin polarization along the initial one
has appeared  when only the electron-LO phonon scattering is included.
This peak is absent when the Coulomb scattering is added.
At $t=35$\ ps, the third peak also appears in the case with
 the Coulomb scattering, but with a smaller magnitude.
Later at $t=60$\ ps, it is interesting to see that the magnitude of
the third peak in the case with the coulomb scattering exceeds the one
without, which indicates a longer spin diffusion
length with the Coulomb scattering.

These features are understood as following: The interchange of the
momentums between the oppositely moving spin-up and -down electrons
due to the Coulomb scattering  induces an effective friction,
which is known as the Coulomb drag.\cite{drag,drag1}
This drag weakens the diffusion
of the spin pulse dramatically as shown in Fig.\ \ref{fig3}. In
addition to the Coulomb drag, the Coulomb scattering also markedly affects
the spin dephasing.\cite{many} It drives electrons to a more homogeneous
states in the {\bf k}-space, suppresses the anisotropy caused by the DP term,
and reduces the spin dephasing.\cite{many} Both effects
slow down the decay of the spin signal at the center of the pulse. Moreover,
the spin diffusion length is determined by
the above two counter effects: On one hand, the Coulomb drag
shortens the spin diffusion length; On the other hand, the increase of
the spin dephasing time by the Coulomb scattering leads to a longer spin
diffusion length. And our results indicate that for the case of
our study, the later effect is
more important in the long run, which gives rise to a longer spin diffusion
length.

We further explore the effect of impurities on spin diffusion in
the presence of the Coulomb scattering. The result is also plotted
in Fig.\ \ref{fig3}. From the figure one can see that
as expected, the spin diffusion becomes much slower due to the
decrease of the mobility. This effect, together with the
well known fact that the spin dephasing time is increased
 by the electron-impurity
scattering,\cite{meier,wu1}  make the decay of the
spin polarization at the center
of the pulse much slower.

 Finally we investigate the transport of the spin pulse under an external
electric field  $E=500$\ V/cm along the $-x$-direction.
The hot-electron effect is not important under this field.\cite{weng2}
Figure \ref{fig4} shows
the  differences of the  spin-up and -down electron densities
versus the diffusion length $x$ at $t=$\ 6, 12, 16 and 24\ ps
in different cases such as  with/without the Coulomb and
impurity scattering. From the figure, one finds that due to the
electric field, the spin signals do not diffuse
symmetrically around the center any more, but transport against the
direction of the electric field. Moreover, the features of the
transport/diffusion of
the  spin polarization in the side against the external
electric field become much richer than the side parallel to the
field. All these results are consistent with what observed in the
experiment.\cite{crooker} From the figure, one can further see the effect
of the Coulomb scattering by comparing the curves  with (solid curve)
and without (dashed curve) the Coulomb
scattering. It is found that similar to the field free case,
the spin signal is more difficult to
diffuse/transport out due to the Coulomb drag. Moreover as the
Coulomb scattering impedes the spin dephasing, combined with the
drag effect, it leads to the slower decay of the spin signal at the
center.   When the impurity with impurity density $N_i=0.5N_e$
is added to the system, it further slows the spin
transport and the decay of the center of the spin
signal as the impurity scattering impedes both the mobility and the
spin dephasing.

It is noted that in our two-dimensional case, the temperature is
much higher than the one in the experiment (4\ K).\cite{crooker}
It is because the system in the experiment is a bulk
one where the effective magnetic field from the DP term and the strain
is as weak as $\sim1.6\times10^{- 3}$\ T. It therefore
requires a much lower temperature to reduce the scattering in order to
observe the reverse of the spin
polarizations. While in QW's with small well width, due to the
strong confinement, the effective magnetic field is about $3.4$\ T
and the phenomenon can sustain a much stronger
scattering at high temperatures.

In conclusion, we have performed a study on the time evolution of spin
pulses at a high temperature (200\ K) through self-consistently
solving the kinetic spin Bloch equations  with the electron-LO phonon,
electron-impurity and electron-electron Coulomb scattering explicitly
 included.
The effect of the Coulomb scattering on the spin diffusion/transport
is clearly demonstrated: The Coulomb
scattering on one hand prolongs the spin dephasing time, which helps to
sustain the spin signal. On the other hand, it causes the
Coulomb drag which weakens the diffusion
of the spin signal. In the case
of our study, the Coulomb scattering
leads to a slower spin diffusion
speed but a longer spin diffusion length.

It is also shown that the Coulomb scattering does not kill the striking
feature of the spin oscillation and the spin reverse along the spin
diffusion direction, which were originally predicted
where only the electron-LO phonon scattering is
considered.\cite{weng_JAP} In addition,
when an external electric field is applied, the spin signal tends to
transport against the direction of the electric field. All these
phenomena have been observed in a recent experiment in bulk
GaAs at very low temperature.\cite{crooker} Our study shows
that one may observe these features at much higher temperature in
QW's with small well width. Experiments are suggested to verify these
effects.

Moreover we also study the effects of the pulse width and the impurity on
the spin diffusion. The results indicate that the wider the pulse
width is, the weaker the spin diffusion and the fewer
the times of the reverse of the spin polarization along the diffusion.
Also the impurity affects the spin diffusion in
two competing directions, {\em i.e.}
impeding the spin diffusion and reducing the spin dephasing.

\bigskip
\bigskip

This work was supported by the Natural Science Foundation of China
under Grant Nos. 90303012 and 10574120, the Natural Science Foundation
of Anhui Province under Grant No. 050460203  and SRFDP.

\end{document}